\documentclass[journal,onecolumn,11pt]{IEEEtran}
\usepackage{amsmath}                
\usepackage{amsthm}                 
\usepackage{amssymb}
\usepackage{wrapfig}
\theoremstyle{plain}

\usepackage[usenames,dvipsnames]{xcolor}
\usepackage[colorlinks=true, citecolor=blue, urlcolor=blue, linkcolor=blue]{hyperref}       
\usepackage[super,compress]{natbib}
\usepackage{multibib}

\newcites{M}{Methods Resources}

\newcommand{\liinesfig}[3]{\renewcommand{\figurename}{Fig.}\begin{figure}[htb!]\begin{center}\includegraphics[width=3.25in]{#1}\vspace{-0.1in}\caption{#2}\label{Fig:#3}\end{center}\end{figure}}
\newcommand{\liinesbigfig}[4]{\begin{figure}[ht!]\begin{center}\includegraphics[width=#4in]{#1}\vspace{-0.1in}\caption{#2}\label{#3}\end{center}\vspace{-0.2in}\end{figure}}

\theoremstyle{definition}
\newtheorem{defn}{Definition}

\ifCLASSINFOpdf
   \usepackage[pdftex]{graphicx}
   \graphicspath{{../pdf/}{../jpeg/}}
   \DeclareGraphicsExtensions{.pdf,.jpeg,.png}
\else
\fi

\ifCLASSOPTIONcompsoc
 \usepackage[caption=false,font=normalsize,labelfont=sf,textfont=sf]{subfig}
\else
 \usepackage[caption=false,font=footnotesize]{subfig}
\fi

\usepackage{dblfloatfix}

\begin{document}
%
\title{Hetero-functional Graph Resilience of the Future American Electric Grid}

\author{
Thompson, D. J.\quad
\and
Schoonenberg, W. C. H.\quad
\and
Farid, A. M.
\\ Thayer School of Engineering at Dartmouth college, Hanover, NH, USA
}

\maketitle




\vspace{-0.5in}

\begin{wrapfigure}{R}{0.4\textwidth}
\vspace{-0.2in}
\begin{center}
\centering \includegraphics[width=0.38\textwidth]{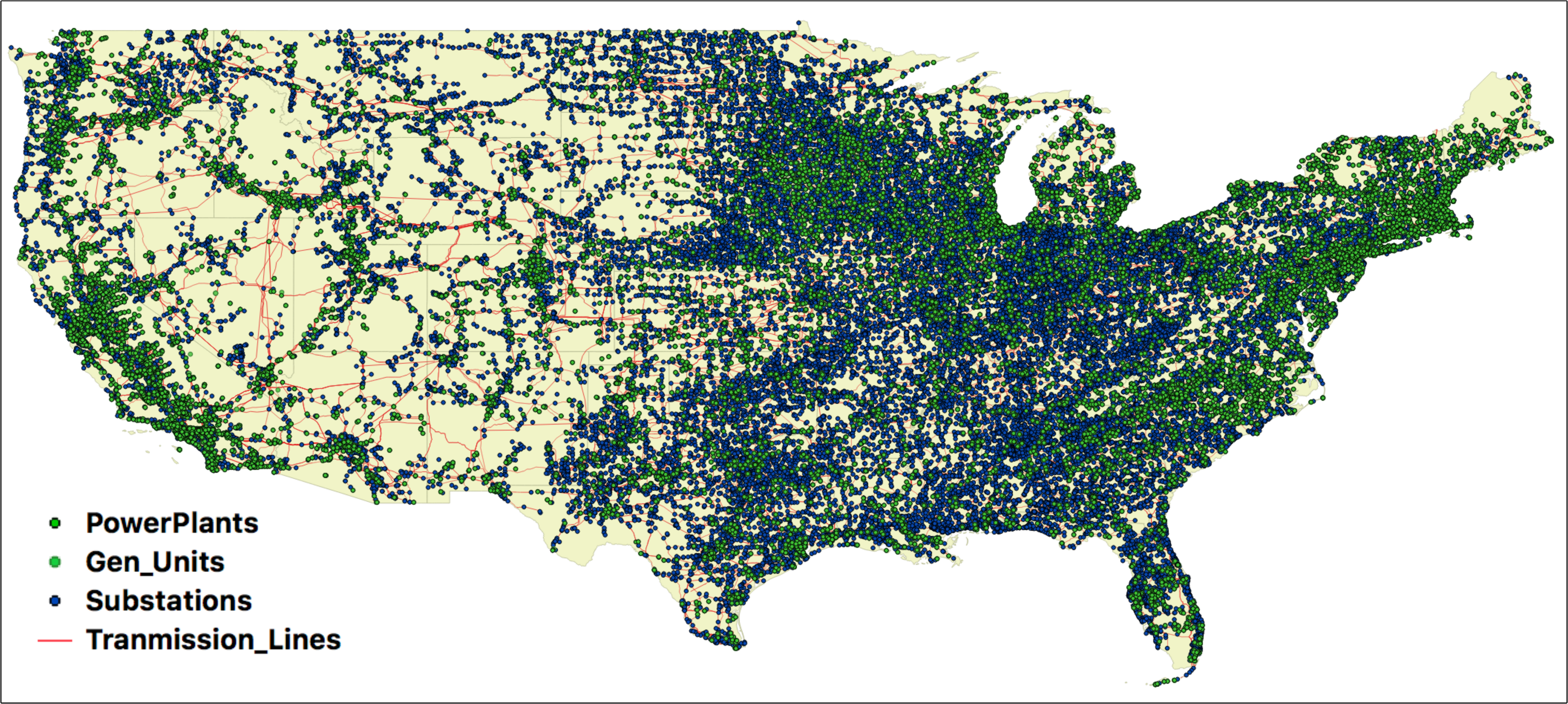}
\end{center}
\vspace{-0.2in}
\caption{A GIS Visualization of the American Electric Power System.}\label{platts_data}
\vspace{-0.2in}
\end{wrapfigure}
\textbf{As climate change takes hold in the 21$^{st}$ century, it places an impetus to decarbonize the American electric power system with renewable energy resources.   There is a broad technical consensus{\color{black}\cite{Muhanji:2018:SPG-J37,Annaswamy:2013:SPG-BK02,Stoustrup:2017:00}} that these renewable energy resources can not be integrated alone but rather require a whole host of profound changes in the electric grid's architecture; including meshed distribution lines, and energy storage solutions.  One question that arises is whether these three types of mitigation measures required by decarbonization will also serve as adaptation measures when the climate changes and extreme weather phenomena become more prevalent.   Consequently, this paper presents a structural resilience analysis of the American electric power system that incrementally incorporates these architectural changes in the future.  Building upon a preliminary study\cite{Thompson:2020:SPG-C68}, the analysis draws on an emerging \textbf{\emph{hetero-functional graph theory}}{\color{black}\cite{Schoonenberg:2018:ISC-BK04}} based upon the inter-connectedness of a system's \textbf{\emph{capabilities}}.  The hetero-functional graph analysis confirms our formal graph understandings from network science\cite{Albert:2000:00,Albert:2004:00,Amaral:2000:00} in terms of cumulative degree distributions and traditional attack vulnerability measures.  The paper goes on to show that hetero-functional graphs relative to formal graphs more precisely describe the changes in functionality associated with the addition of distributed generation and energy storage as the grid evolves to a decarbonized architecture.  Finally, it demonstrates that the addition of all three types of mitigation measures enhance the grid's structural resilience; even in the presence of disruptive random and targeted attacks.  The paper concludes that there is no structural trade-off between grid sustainability and resilience enhancements and that these strategic goals can be pursued simultaneously.}   

\begin{figure}[t]
\centering
\includegraphics[width=6.5in]{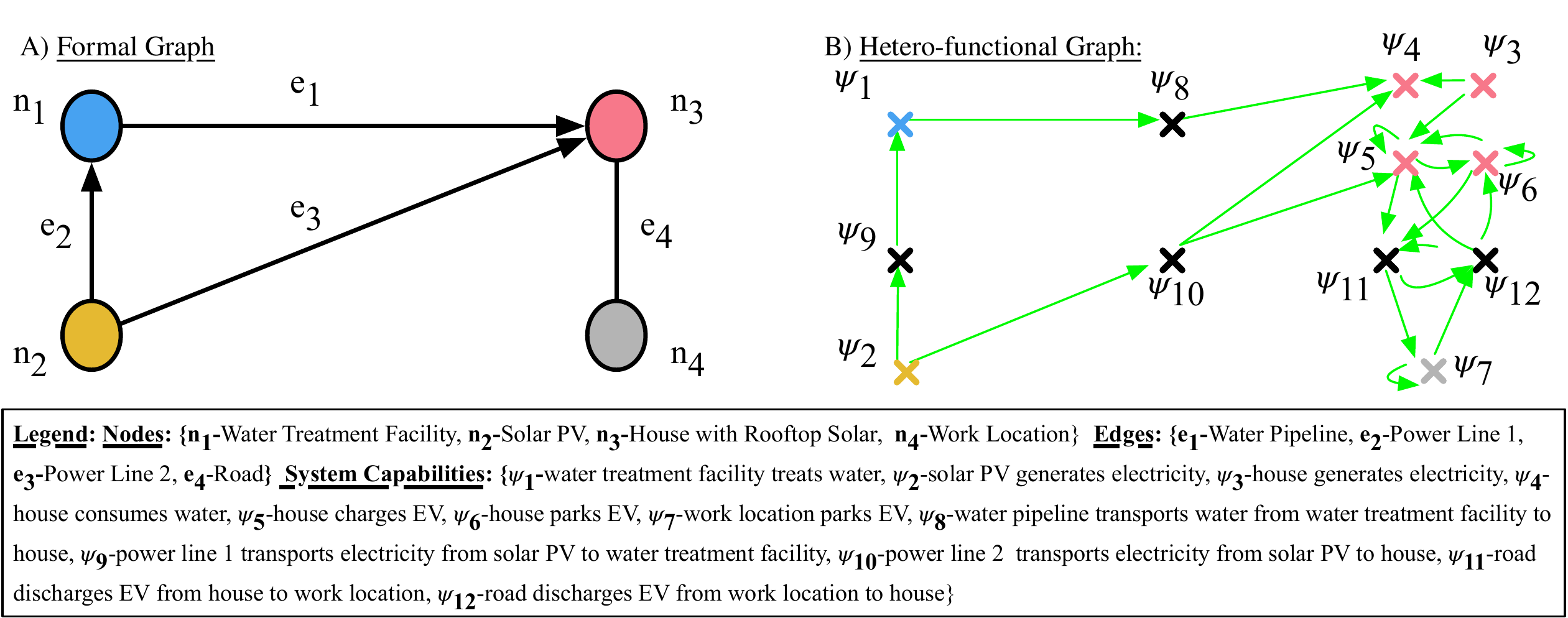}
\vspace{-0.1in}
\caption{A Visual Comparison of a Formal Graph (FG) and a Hetero-functional Graph (HFG) Model of the Same Hypothetical System.}\label{TG_vs_HFGT}
\vspace{-0.3in}
\end{figure}

While renewable energy can be integrated centrally at a utility-scale, one of its primary benefits is that it can empower end-consumers with distributed generation (DG) in the form of rooftop solar photovoltaics (PV), small-scale wind turbines, and even run-of-river hydro power.  The potential for power back-flow in an electric distribution system designed for one-way outward flow requires a migration from a radial to a meshed topology.   In the meantime, these DG resources are intermittent and often require complementing energy storage solutions.  These three additions represent fundamental changes to the system architecture of the American Electric Power System (AEPS).  

Such architectural changes can have a profound impact on the system's \textbf{\emph{resilience}} in terms of its own ability to withstand disruptions; be they natural, artificial, or intentionally nefarious\cite{Albert:2000:00,Albert:2004:00,Amaral:2000:00}.  To address such questions quantitatively, the network science community has used graphs to mathematically represent the \textbf{\emph{form}} of a system\cite{Crawley:2015:00}.  The graph's nodes are made to correspond to the grid's power plants, substations, and consumers while the graph's edges are made to correspond to the grid's power lines.  For clarity, we refer to such a mathematical model as a \textbf{\emph{formal}} graph (FG)\cite{Crawley:2015:00}.  Consequently, system resilience can be quantitatively studied in terms of successive node or edge failures.  While such a simple graph model can address the resilience improvements caused by a migration towards meshed distribution networks, and the addition of \textbf{\emph{new nodes}} that represent solar PV and energy storage solutions, it is ill-equipped to address the integration of such distributed energy resources on \textbf\emph{existing} nodes as in the case of solar panels on rooftops and batteries at homes, substations, and centralized generators.  In effect, such additions (as shown later) do not numerically change the formal graph, and consequently, have no effect on the value of the associated resilience measure.

In contrast, the model-based systems engineering (MBSE) community recognizes that a formal graph representing system form is merely a subset of system architecture and that a more comprehensive description of architecture must also describe:  1.)  a set of functions that the system performs and 2.) the allocation of those functions to the elements of form\cite{Crawley:2015:00,Friedenthal:2011:00}.  While the MBSE literature normally describes system architecture using graphical models (e.g. UML \& SysML)\cite{Friedenthal:2011:00}, hetero-functional graph theory has developed to translate these models into their quantitative equivalents\cite{Schoonenberg:2018:ISC-BK04}.  Consequently, hetero-functional graphs are able to explicitly and quantitatively describe the incorporation of new functionality onto existing formal nodes as in the case of rooftop solar and home batteries.

The simple example in Fig. {\color{black}\ref{TG_vs_HFGT}} illustrates the differences between a formal graph and a hetero-functional graph (HFG).  The formal  graph (in Fig {\color{black}\ref{TG_vs_HFGT}}a.) shows a system composed of four nodes: a water treatment facility, a solar PV panel, a house with rooftop solar, and a work location.  These are connected by four edges:  a water pipeline, two power lines and two roads.  Fig {\color{black}\ref{TG_vs_HFGT}}b. shows the associated hetero-functional graph following the methods described in the methods section.  Instead of four nodes that represent point-like facilities, the hetero-functional graph now has 12 nodes that represent \textbf{\emph{system capabilities}}.  The water treatment facility, solar PV panel, and work location appear unchanged between the two graphs because they each have only one capability.  In contrast, the house with rooftop solar provides four capabilities in the HFG.  This multiplicity of capabilities assigned to a single facility forms the basis upon which to investigate the effect of DG on electric power system resilience.  Thirdly, the edges in the formal graph now appear as transportation capabilities (\textbf{\emph{nodes}}) in the HFG.  Finally, the directed edges in the HFG indicate the logical sequences of these capabilities such that if one were to follow them a ``story" of capabilities would emerge.  (i.e. The water treatment facility treats water $(\psi_1)$ and then the water pipeline transports the water from the water treatment facility to the house $(\psi_8)$).  It is important to recognize that because the FG and HFG have different quantities of nodes and edges, the associated values of graph measures also differ. 

\begin{wrapfigure}{R}{0.3\textwidth}
\vspace{-0.2in}
\begin{center}
\centering \includegraphics[width=0.28\textwidth]{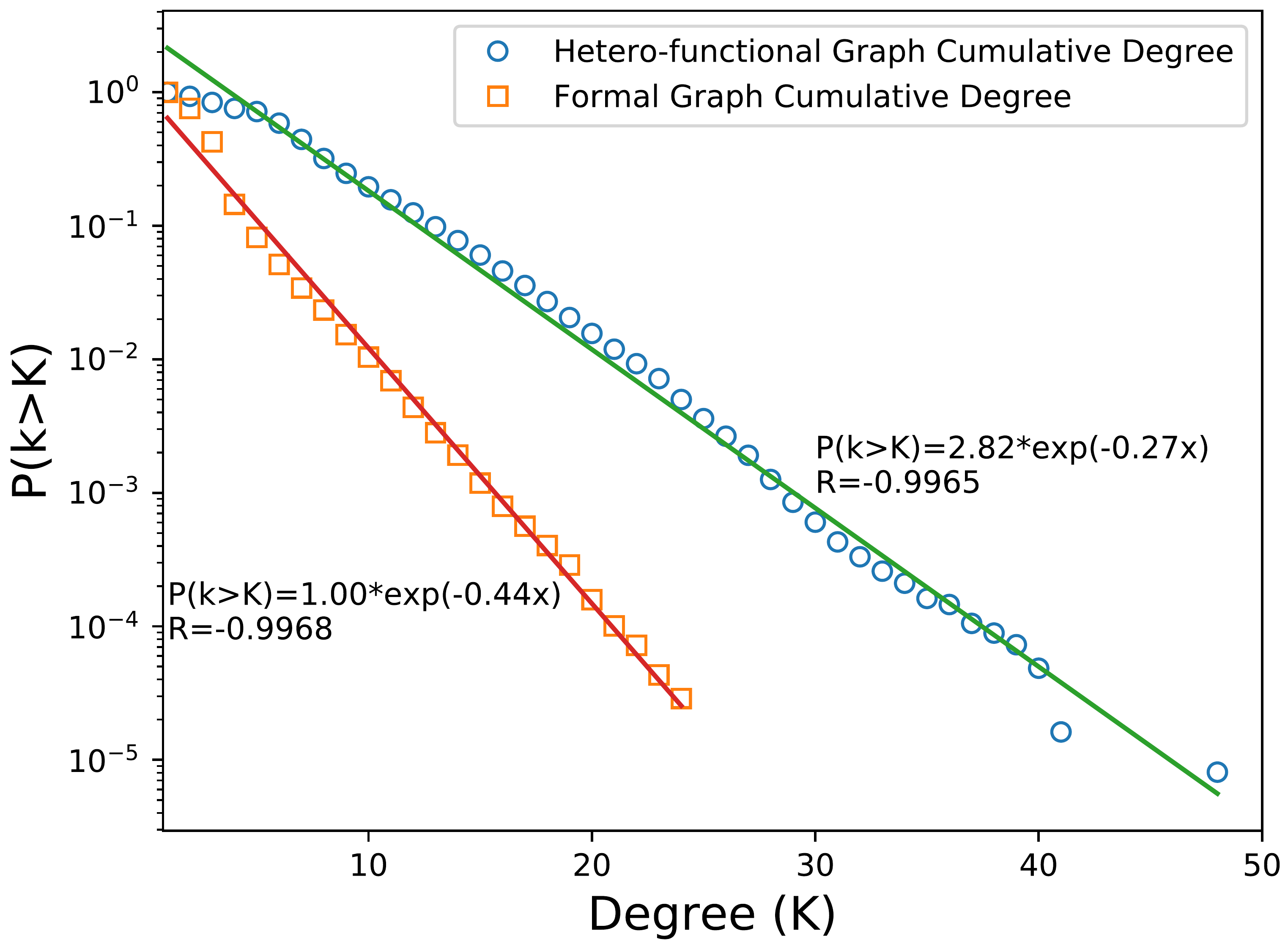}
\end{center}
\vspace{-0.2in}
\caption{The formal and hetero-functional graphs of the AEPS have an exponential-tail cumulative degree distribution indicating a single-scale small-world network.}\label{degree_dist}
\vspace{-0.2in}
\end{wrapfigure}
Despite these apparent differences, FGs and HFGs, when studied in the context of the AEPS,  demonstrate remarkable similarities in their degree distributions.  The Platts Map Data Pro data set\cite{Platts:2017:00} was used to conduct the analysis.  As shown in Fig. \ref{platts_data}, it consists of a GIS layer with {\color{black}13,568} power plants, {\color{black}34,649} generation units, {\color{black}78,880} substations, and {\color{black}104,329} transmission lines.  A FG graph adjacency matrix is readily extracted from this GIS data and the methods section describes the construction of the associated HFG.  Both the FG and HFG, as shown in Fig.  \ref{degree_dist}, confirm the network science result\cite{Amaral:2000:00,Albert:2004:00} of a cumulative degree distribution with exponential decay law $P(k\leq K)c e^{-\alpha k}$.  They have exponential coefficients of {\color{black}  $\alpha_{FG}=0.44$} and {\color{black}$\alpha_{HFG}=0.27$} respectively.  This result suggests that the underlying single-scale small-world socio-technical dynamic of preferential attachment of transmission lines appears in both network models.  The larger exponential coefficient in the HFG arises because it always has more nodes and edges than its FG counterpart.  Nevertheless, the presence of small-world structure in both graphs indicates similar behavior with respect to resilience and attack vulnerability.  

\liinesbigfig{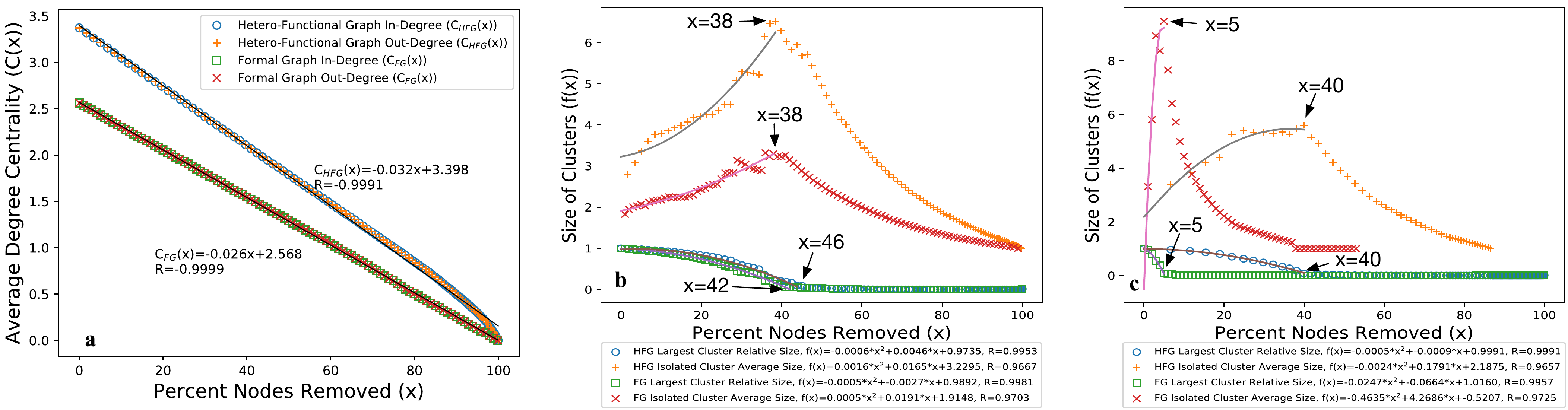}{The FG and the HFG of the AEPS demonstrate similar behavior in: \textbf{a} their average degree centrality when subjected to successive random attacks, \textbf{b} the size of their largest cluster and average size of their isolated clusters when subjected to random attacks, and \textbf{c} the size of their largest cluster and average size of their isolated clusters when subjected to targeted attacks.}{TG_vs_HF_cent_size}{6}

In order to verify this hypothesis, the FG and HFG models of the AEPS were subjected to nodal attacks and then assessed with respect to their average degree centrality, size of largest cluster, and average size of isolated cluster.  (See Supp. Materials for details).  Fig. \ref{TG_vs_HF_cent_size}a shows the random attack vulnerability of the FG and HFG with respect to in-degree and out-degree centrality measures.  The parity of in-degree and out-degree centrality  in the case of the FG is caused by its undirected nature, while in the case of the HFG, it is caused by the two transportation capabilities assigned to each power line combined with the single capability assigned to power plants and substations.  Both the FG and HFG show that the average degree centrality degrades linearly with the fraction of randomly removed nodes.  The slopes of their best-fit lines are {\color{black}-0.026 and -0.032} respectively, and their regression coefficients are {\color{black}-0.9999 and -0.9991} respectively.  This intuitive phenomena occurs because random attacks do not discriminate on the basis of the nodal-degree.  Consequently, the successive removal of two sets of nodes do not change the average number of edges lost.  Fig. \ref{TG_vs_HF_cent_size}b confirms the network science result{\color{black}\cite{Albert:2000:00}} that the size of the largest cluster of the FG of an electric power system degrades quadratically under random attacks before reaching an inflection point (at {\color{black}x=42\%} in this case).  Fig. \ref{TG_vs_HF_cent_size}b also confirms that the average size of isolated clusters first grows to a relative peak (at {\color{black}x=38\%} in this case) before sharply falling again.  These two behaviors also appear in the associated HFG with a largest cluster inflection point of {\color{black}x=46\%} and a peak average size of isolated cluster at {\color{black}x=38\%}.  Fig. \ref{TG_vs_HF_cent_size}c also shows that the FG and HFG confirm the results of the published literature{\color{black}\cite{Albert:2000:00}} regarding the structural response to targeted attacks.  This time, the size of the largest cluster reaches an inflection point at {\color{black}5\%} and {\color{black}40\%} for the FG and HFG respectively.  Also, the peak average size of the isolated cluster occurs at {\color{black}5\%} and {\color{black}40\%} respectively.  In short, all four of these structural responses to nodal attacks in the FG are matched closely in the associated HFG.   The differences in the labelled critical points stem from the larger number of nodes and edges in the HFG.  Collectively, the results in Fig. \ref{TG_vs_HF_cent_size} show that the FG and HFG can be used interchangeably to study traditional attack vulnerability measures.  

\vspace{-0.15in}
\liinesbigfig{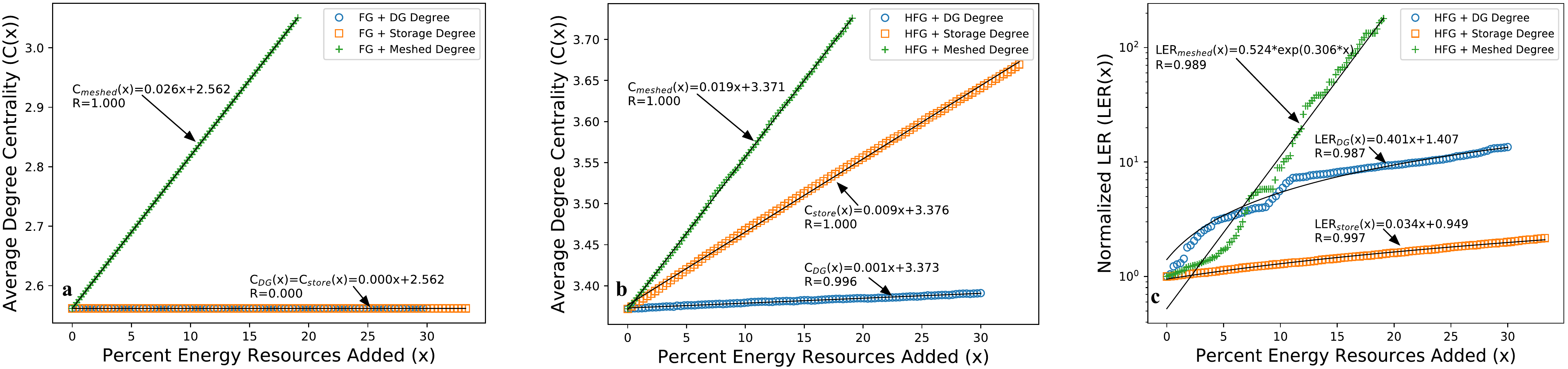}{The FG \textbf{a} and the HFG \textbf{b} of the AEPS demonstrate differing behavior in: their average degree centrality when subjected to successively adding architectural and functional improvements. The FG \textbf{a} is only able to capture the improvement of adding meshed power lines while the HFG \textbf{b} measure improvements from the additions of distributed generation and storage as well.  The Latent Engineering Resilience of the HFG\textbf{c} measures the potential improvements from adding all three forms of additions to the AEPS.}{add_cent_fg_hfg}{6}

Such a conclusion serves as the foundation upon which to investigate the AEPS as it migrates towards a decarbonized system architecture.  More specifically, and as detailed in the methods section, the AEPS was subjected to incremental additions of distributed generation, energy storage resources, and meshed power distribution lines.   While the last of these is clearly a change in the AEPS formal topology, the first two represent fundamental changes in the AEPS' \textbf{\emph{functionality}} without a commensurate change in formal topology.  Indeed, from the perspective of an electric power utility or grid operator, the FG does not change when end-users add rooftop solar and batteries to their buildings.  Fig. \ref{add_cent_fg_hfg}a shows the effect of adding these resources on the average degree centrality of the FG.  As expected, the addition of new meshed power distribution lines increases the average degree centrality linearly with a slope of {\color{black}0.026} and a regression coefficient of 1.  In contrast, the addition of distributed generation and energy storage resources has \textbf{\emph{no effect}} on the average degree centrality of the FG because the underlying adjacency matrix remains \textbf{\emph{entirely unchanged}}.  
Such a result calls into question either the adequacy of the FG as a \textbf{\emph{model}} or the adequacy of degree centrality as a resilience \textbf{\emph{measure}}.  After all, an end-user with newly installed distributed generation or energy storage would continue to have some form  of electric power service even if they were entirely disconnected from the rest of the grid.  

In contrast, Fig. \ref{add_cent_fg_hfg}b shows that the degree centrality of the HFG responds as the AEPS migrates towards a decarbonized system architecture.  Again, as expected, the addition of new meshed power distribution lines increases the average degree centrality linearly with a slope of {\color{black}0.019} and a regression coefficient of {\color{black}one}.  Furthermore, the introduction of energy storage and distributed generation resources now increases the average degree centrality linearly with a slope of {\color{black}0.009} and {\color{black}0.001} respectively and a regression coefficient of {\color{black}one} and {\color{black}0.996} respectively.  Although, the topology of the underlying formal graph experiences no change during the addition of these two types of resources, the HFG incorporates both new capability nodes as well as connecting edges.  More specifically, each new distributed generation resource adds a new capability node and a new edge.   In the meantime, each storage resource adds a new capability node but adds at least four new edges.  These results show that the HFG is more adequate than the FG as a model when system architecture is changing its functionality and not just its formal topology.  

The results of Fig. \ref{add_cent_fg_hfg}b, however, understate the ``resilience value" of distributed generation.  Again, from a practical perspective,  an end-user would not differentiate between electricity supplied from the grid or that supplied from distributed generation.   In contrast, a degree centrality measure only values the resilience that comes comes from greater grid connectivity.  As an alternative,  Fig. \ref{add_cent_fg_hfg}c calculates ``latent engineering resilience" (LER)\cite{Farid:2015:ISC-J19} measure as the AEPS responds to the same architectural changes found in Fig. \ref{add_cent_fg_hfg}b.  As explained in the methods section, the LER measures was specifically developed to calculate the number of viable service paths in hetero-functional graphs.   As expected, Fig. \ref{add_cent_fg_hfg}c shows that the LER grows exponentially (with a coefficient of {\color{black}$\alpha=0.306$ and regression coefficient of 0.989)} as meshed distribution lines are added.  Such lines will exponentially increase the number of available service paths.  In the meantime, the LER grows linearly with addition of DG with a slope of {\color{black}0.401 and a regression coefficient of 0.987}.  Each new DG resource makes use of the network topology to introduce a relatively large but proportional number of service paths.  Finally, the LER grows linearly 
with a slope of {\color{black}0.034 and a regression coefficient of 0.997} as energy storage resources are added.  Unlike DG, energy storage resources still require generation resources in order to contribute a service path and so their resilience enhancing effect is contingent upon generation and distribution capabilities.  The results in Fig. \ref{add_cent_fg_hfg}  show that the HFG relative to a FG more precisely describes the future evolution of the AEPS' architecture.  Furthermore, the quantification of these resilience improvements is more accurately measured using a LER measure based upon service paths than simply a network centrality measure.   Finally, and most importantly, the addition of DG and energy storage resources in combination with meshed distribution lines enhance the AEPS' transition to a highly resilient and decarbonized system architecture.  In other words, from an architectural perspective, there is no trade-off between grid sustainability and resilience enhancements and that these strategic goals can be pursued simultaneously.   

\liinesbigfig{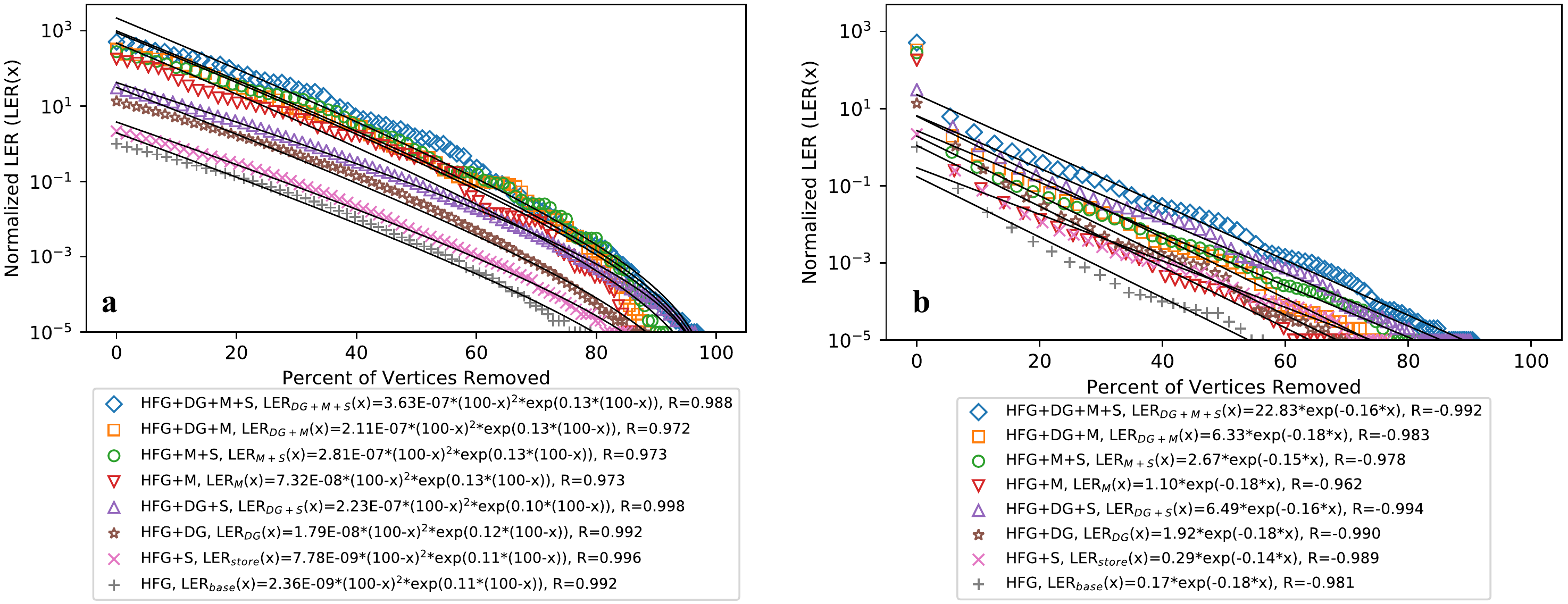}{Eight increasingly sustainable grid architectures are studied in terms of their resilience to random and targeted attacks.   These include all combinations of 33\% of generators with additional storage (S), 30\% of substations with additional distributed generation (DG), and 20\% additional meshed distribution lines (M).  \textbf{a} shows the response to random attacks. \textbf{b} shows the response to targeted attacks based upon highest degree centrality.}{HFG_LER_attack}{6.5}

This conclusion is further investigated in Fig. \ref{HFG_LER_attack}.  More specifically, the LER of eight increasingly decarbonized architectures of the AEPS is studied in presence of random and targeted attacks.  As detailed in the methods section, the eight investigated architectures include all combinations of: 30\% of substations with additional distributed generation (DG), 33\% of generators with additional storage (S), and 20\% additional meshed distribution lines (M).  Irrespective of the choice of decarbonized architecture, these systems respond differently to random and targeted attacks.  In the case of random attacks (Fig. \ref{HFG_LER_attack}a), the LER $\propto (100-x)^2*e^{-\alpha (100-x)}$ while in the case of targeted attacks (Fig. \ref{HFG_LER_attack}b), the LER more closely follows $e^{(-\alpha x)}$.  Targeted attacks on the basis of highest degree centrality remove an exponentially decreasing number of edges as can be inferred from Fig. \ref{degree_dist}.  Each of these edges, in turn, contribute to an exponential number of paths; resulting in an overall exponential effect.   In contrast, the random attacks combine the exponential loss of paths with the parabolicly decreasing size of the largest cluster (shown in Fig. \ref{TG_vs_HF_cent_size}b).  Consequently, and as expected, successive targeted attacks more effectively diminish the grid's LER than random attacks do.  In both cases, the LER measure is able to precisely differentiate between all three types of architectural changes.  In agreement with the results from Fig. \ref{add_cent_fg_hfg}c, the architectures with meshed distribution lines, as a group, exhibit the greatest resilience.  This group is followed by the architectures with distributed generation which is in turn followed by the architectures with energy storage.  All of these cases report higher resilience values than the baseline system representing the AEPS in its present form.  In other words, these results confirm that the evolution of the AEPS' to a decarbonized architecture composed of distributed generation, energy storage, and meshed distribution lines will simultaneously enhance its resilience.  

This paper has presented a structural resilience analysis of the American electric power system as it evolves towards a decarbonized architecture consisting of distributed generation, energy storage, and meshed distribution lines.   To conduct the analysis, it relied on hetero-functional graphs which were shown to confirm our formal graph understandings from network science  \cite{Albert:2000:00,Albert:2004:00,Amaral:2000:00} in terms of cumulative degree distributions and traditional attack vulnerability measures.  Such hetero-functional graphs more precisely describe the changes in functionality associated with the addition of distributed generation and energy storage as the grid evolves to a decarbonized architecture.  Finally, it demonstrates that the addition of all three types of mitigation measures enhance the grid's structural resilience; even in the presence of disruptive random and targeted attacks.  Consequently, there is no structural trade-off between grid sustainability and resilience enhancements and that these strategic goals can be pursued simultaneously.       

\vspace{-0.2in}
{\footnotesize
\bibliographystyle{IEEEtran}
\bibliography{0-AEPS_Resil.bib}}

\newpage
\section{\textbf{Methods}}

\subsection{\textbf{Constructing A Hetero-functional Graph}}
A hetero-functional graph, like any other graph, is constructed by identifying a set of nodes and connecting them with edges.  This is a two step process that we describe simply here and illustrate using the example provided in Fig. \ref{TG_vs_HFGT}.  The interested reader is referred to the hetero-functional graph theory text{\color{black}\cite{Schoonenberg:2018:ISC-BK04}} for a more elaborate exposition.  

Unlike formal graphs where the nodes represent elements of form like power plants, substations, and consumers, the nodes in a hetero-functional graph are \emph{system capabilities}.  A capability is the feasible allocation of a given \emph{system process} (or function) to a given \emph{system resource} as an element of form.  The system processes $P$ describe the functionality of the system in terms of a transitive verb followed by an operand (e.g. ``generate + electric power").  These processes are often classified as (in-place) transformation processes $P_\mu$ and transportation processes $P_\eta$.  The system resources $R$ describe the formal composition of the system in terms of nouns.   For example, in Fig. \ref{TG_vs_HFGT}, the set of system resources $R=\{$Water Treatment Facility,  Solar PV, House with Rooftop Solar, Work Location, Water Pipeline, Power Line 1, Power Line 2, Road$\}$.  (Note that $R=V \cup E$ where V and E are the nodes and edges of a formal graph.)  In the meantime, the system processes $P=\{$treat water, generate electricity, consume water, charge EV, store EV, transport water from water treatment facility to house, transport power from solar PV to water treatment facility, transport power from solar PV to house, discharge EV from house to work location, discharge EV from work location to house$\}$.  The feasible allocation of system process to system form is captured in a \emph{system knowledge base} $J_S$. 

\begin{defn}
\textbf{System knowledge base}:  A binary matrix $J_S$ of size $\sigma(P)\times\sigma(R)$ whose element $J_S(w,v)\in \{0,1\}$ is equal to one when $e_{wv}\in {\cal E}$ (in the SysML sense) exists as a system process $p_w\in P$ being executed by a resource $r_v\in R$.  The $\sigma()$ operator returns the size of a set.  
\end{defn}

\liinesfig{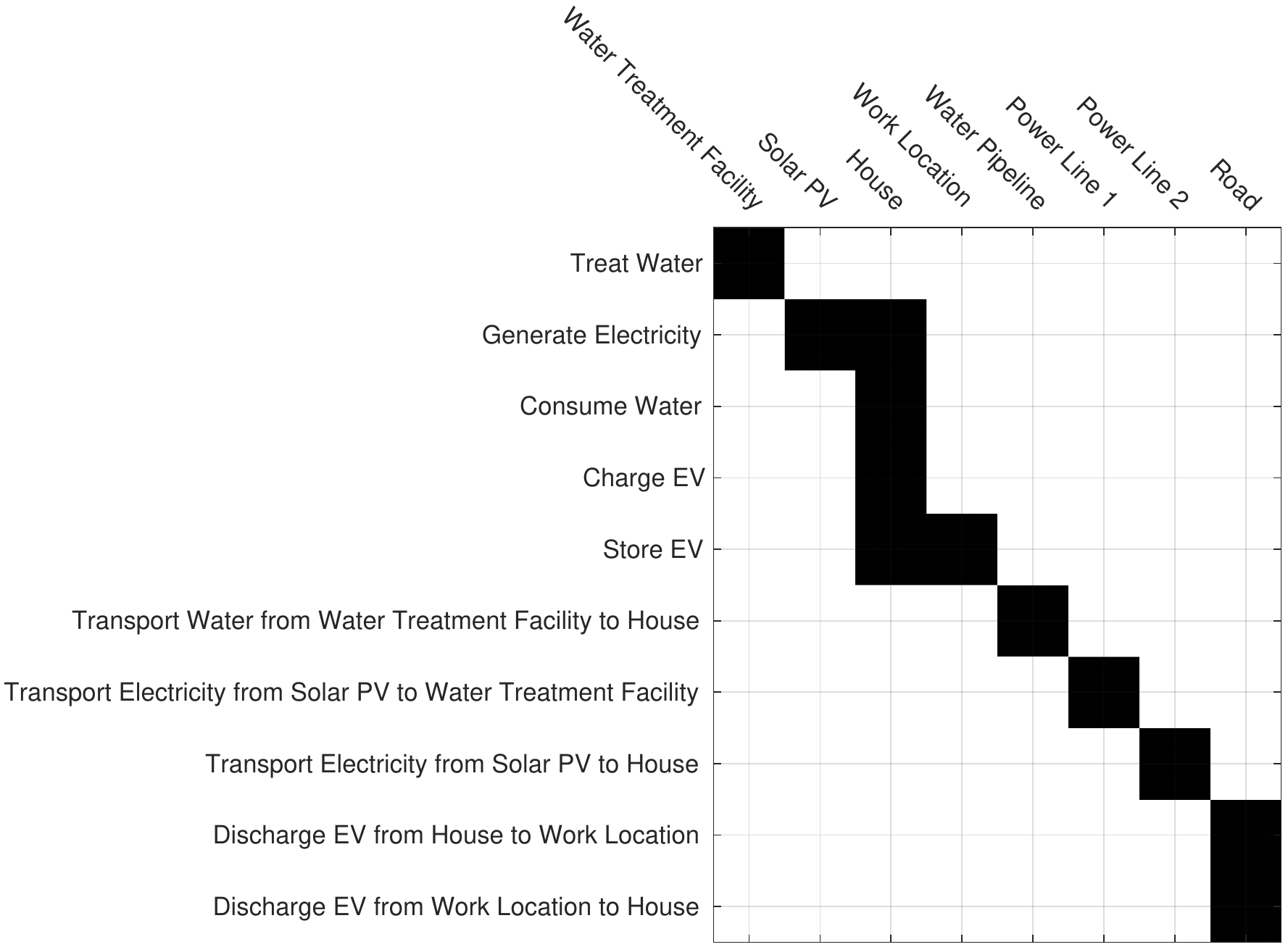}{The hetero functional system Knowledge Base for System Depicted in Fig. \ref{TG_vs_HFGT}.}{FGvsHFG_js}

\noindent Note that the system knowledge base itself constructs a bipartite graph between $P$ and $R$.  The system knowledge base associated with the system depicted in Fig. \ref{TG_vs_HFGT} is shown in Fig. \ref{Fig:FGvsHFG_js} as a monochrome image.  The capabilities of a given resources are explicitly reflected in the associated column of $J_S$.  Consequently, the set of capabilities ${\cal E}=\{$1.) water treatment facility treats water, 2.) solar PV generates electricity, 3.) house consumes water, 4.) house generates electricity, 5.) house charges EV, 6.) house stores EV, 7.) work location stores EV, 8.) water pipeline 1 transports water from water treatment facility to house, 9.) power line 1 transports electricity from solar PV to water treatment facility, 10.) power line 2 transports electricity from solar PV to house, 11.) road ``discharges" EV from house to work location, 12.) road ``discharge" EV from work location to house$\}$.  These capabilities make up the nodes of the hetero-functional graph.  

Once the capabilities of the system have been identified as the nodes of a hetero-functional graph, they can be connected with edges so as to form a hetero-functional adjacency matrix $A_\rho$.    

\begin{equation}\label{A_rho}
    A_\rho=(J_S\ominus K_S)^V(J_S\ominus K_S)^{VT})\ominus K_\rho
\end{equation}
where $()^V$ is shorthand for $vec()$, $\ominus$ denotes Boolean subtraction of matrices, $K_{S}$ is the \emph{system constraints matrix}, and $K_{\rho}$ is the \emph{system sequence constraints matrix}.  $K_S=\textbf{0}$ when all capabilities are functional.     

\begin{defn}
\textbf{System sequence constraints matrix}:  a square binary constraints matrix $K_\rho$ of size $\sigma(R)\sigma(P)\times\sigma(R)\sigma(P)$ whose elements $K(\chi_1,\chi_2)\in\{0,1\}$ are equal to one when string $z_{\chi_1\chi_2}=e_{w_1v_1}e_{w_2v_2} \in Z$ is eliminated.
\end{defn}

\noindent $K_\rho$ is calculated by identifying constraints that impede the logical sequence $z_{\chi_1\chi_2}$ between an ordered pair of capabilities $e_{w_1v_1}e_{w_2v_2}$.  Five types of constraints are possible on $e_{w_1v_1}e_{w_2v_2}$:
\begin{enumerate}
    \item When $w_1$ and $w_2$ indicate transformation processes but $v_1\neq v_2$.
    \item When $w_1$ indicates a transformation process and $w_2$ indicates a transportation process but $v_1$ is not the origin of $w_2$ as a transportation process.  
    \item When $w_1$ indicates a transportation process and $w_2$ indicates a transformation process but $v_2$ is not the destination of $w_1$ as a transportation process.  
    \item When $w_1$ and $w_2$ indicate  transportation processes but the destination of the former is not equivalent to the origin of the later.  
    \item When $w_1$ and $w_2$ are not permitted by the functional reference architecture of the system.  For example, in electric power systems, the generation of electric power systems is followed by any number of transportation processes which is followed by the consumption of electric power.
\end{enumerate}
Finally, it is often useful to use a projection operator $\mathbb{P}$ to eliminate the empty rows and columns in $A_\rho$.  
\begin{equation}
    \tilde{A}_\rho=\mathbb{P}{A}_\rho\mathbb{P}^T
\end{equation}
From these steps, the hetero-functional adjacency matrix $\tilde{A}_\rho$ corresponding to Fig. {\color{black}\ref{TG_vs_HFGT}} is shown in Fig. {\color{black}\ref{TG_vs_HFG_ARP}} below.  It is contrasted with the associated formal graph in Fig. \ref{TG_vs_HFG_ARF}.

\begin{figure}[ht]
\centering
\subfloat{\includegraphics[width=3in]{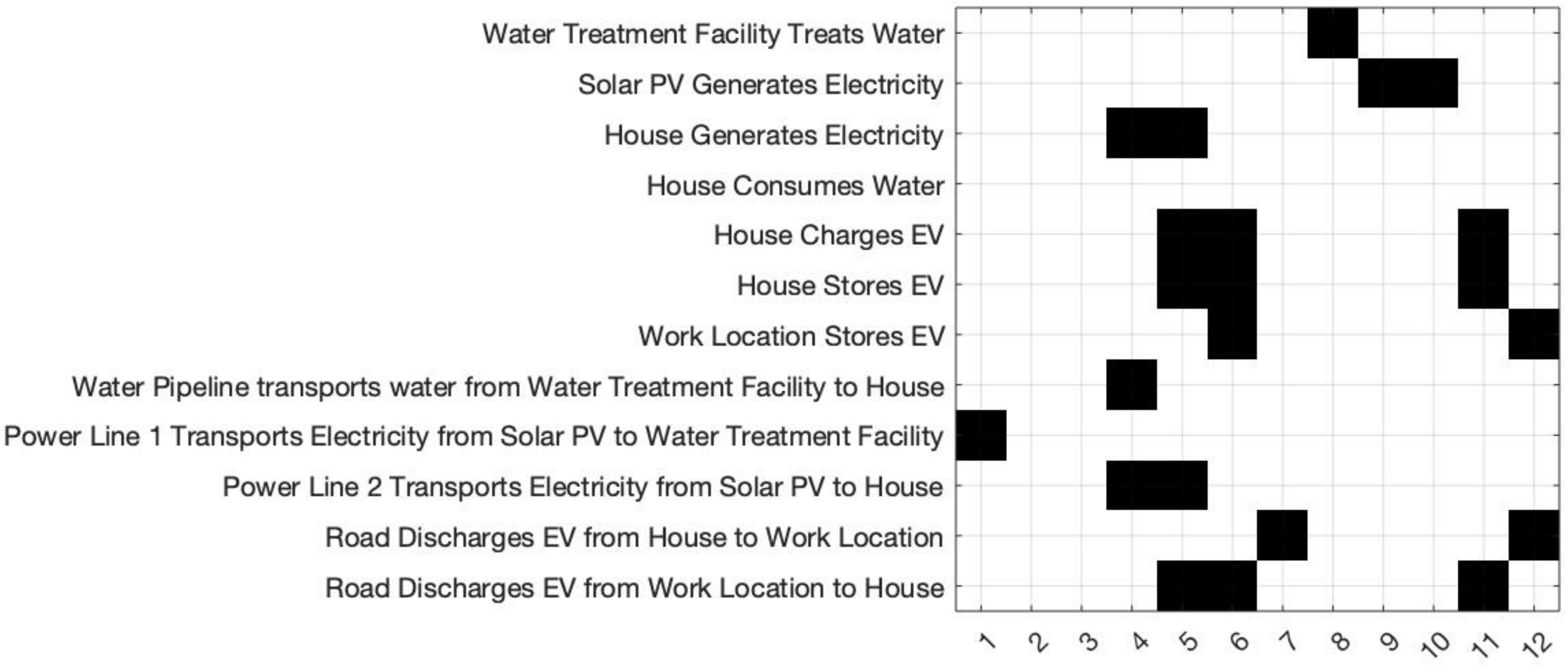}\label{TG_vs_HFG_ARP}}
\qquad
\subfloat{\includegraphics[width=1.74in]{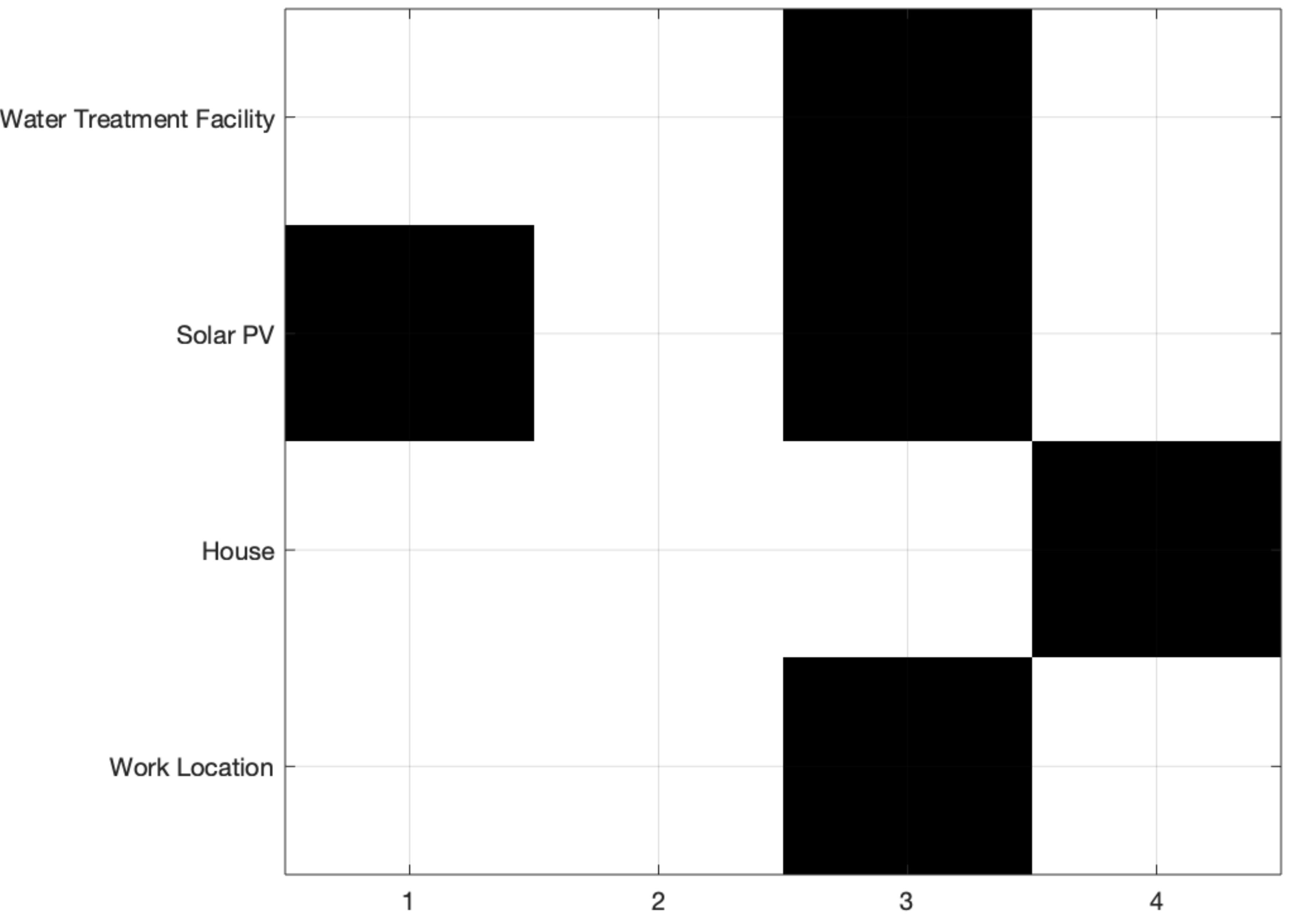}\label{TG_vs_HFG_ARF}}
\caption{The (Projected) Hetero-functional Graph Adjacency Matrix \textbf{(a)} of the System Depicted in Fig. \ref{TG_vs_HFGT}.  The Formal Adjacency Matrix \textbf{(b)} of the System Depicted in Fig.  \ref{TG_vs_HFGT}.}
\label{TG_vs_HF_AR}
\end{figure}

\subsection{\textbf{Constructing a Hetero-functional Graph of the American Electric Power System}}
The Platts Map Data Pro data was used to create the hetero-functional graph of the American electric power system in three steps:

\begin{enumerate}
    \item The Platts Map Data Pro data was converted into a corresponding KML file.  
    \item The KML file was processed into an XML file compatible with the  hetero-functional graph theory toolbox{\color{black}\citeM{Hegde:2020:ISC-JR03}}.  
    \item The hetero-functional graph theory toolbox provides the functionality to automatically calculate the hetero-functional adjacency matrix $\tilde{A}_\rho$.  
\end{enumerate}

The Platts Map Data Pro is a proprietary Geographic Information Systems (GIS) database that contains 38 different GIS data sets of North American energy infrastructure and its associated markets.  This paper focuses on the transmission system and thus uses four GIS layers to create the hetero-functional graph:  the 1.) Power plants, 2.) Generation units (i.e. individual generation facilities within power plants), 3.) Substations, and 4.) Transmission Lines.  Each of these North American layers were cropped to return elements from the United States' electric grid.  This GIS data can be straightforwardly exported as a KML file using the in-built functionality of a capable GIS editor (e.g. QGIS as an open-source solution, and ArcGIS as a leading commercial software).  

\begin{figure}[ht]
\centering
\subfloat{\includegraphics[width=5in]{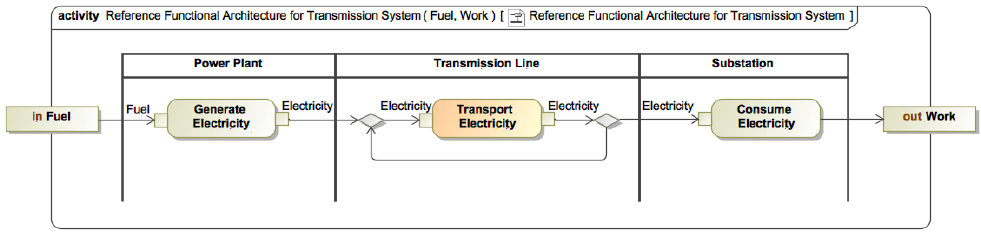}\label{ex_hfg_act}}
\qquad
\subfloat{\includegraphics[width=3in]{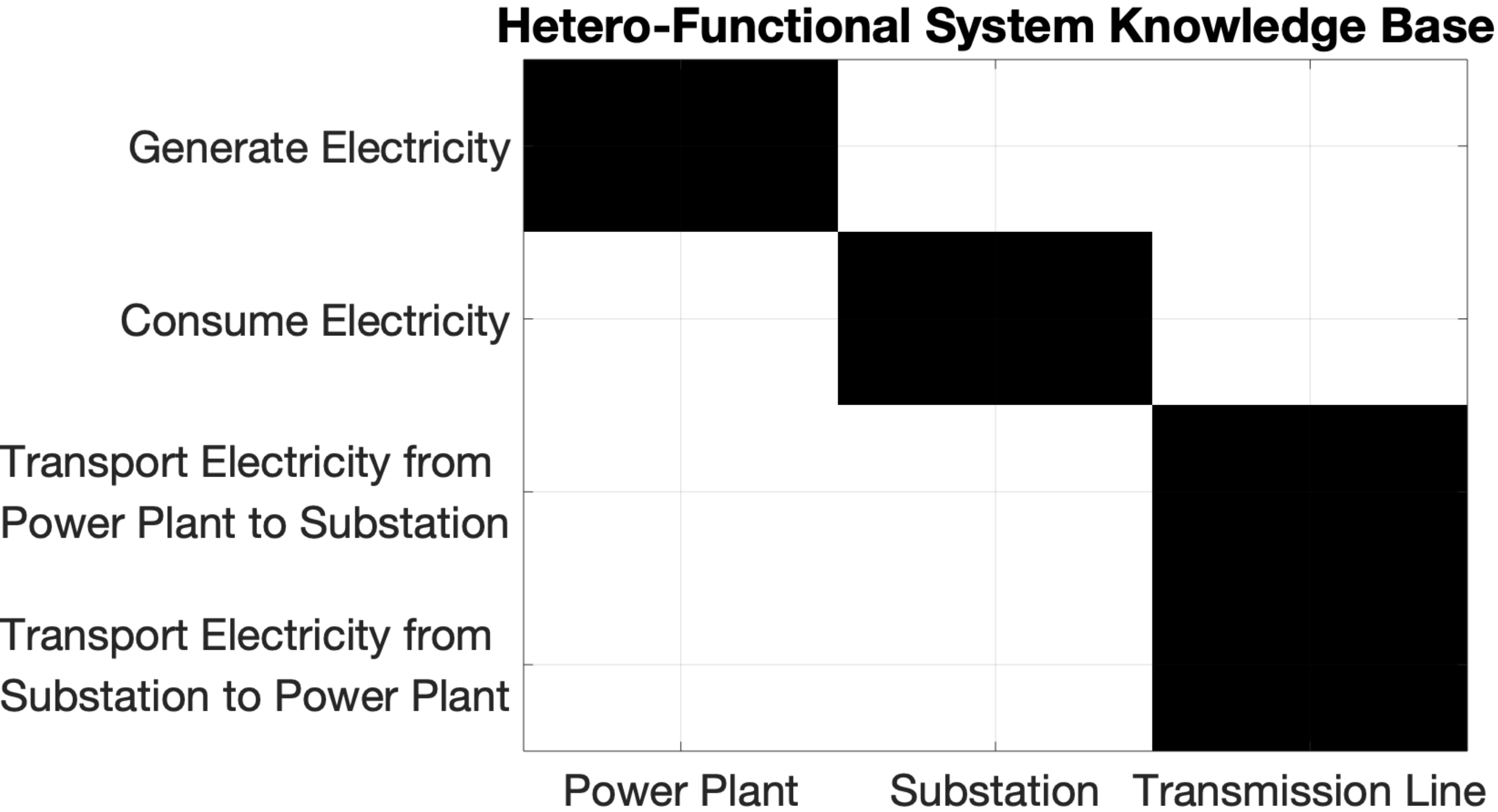}\label{ex_JS}}
\quad
\subfloat{\includegraphics[width=2.78in]{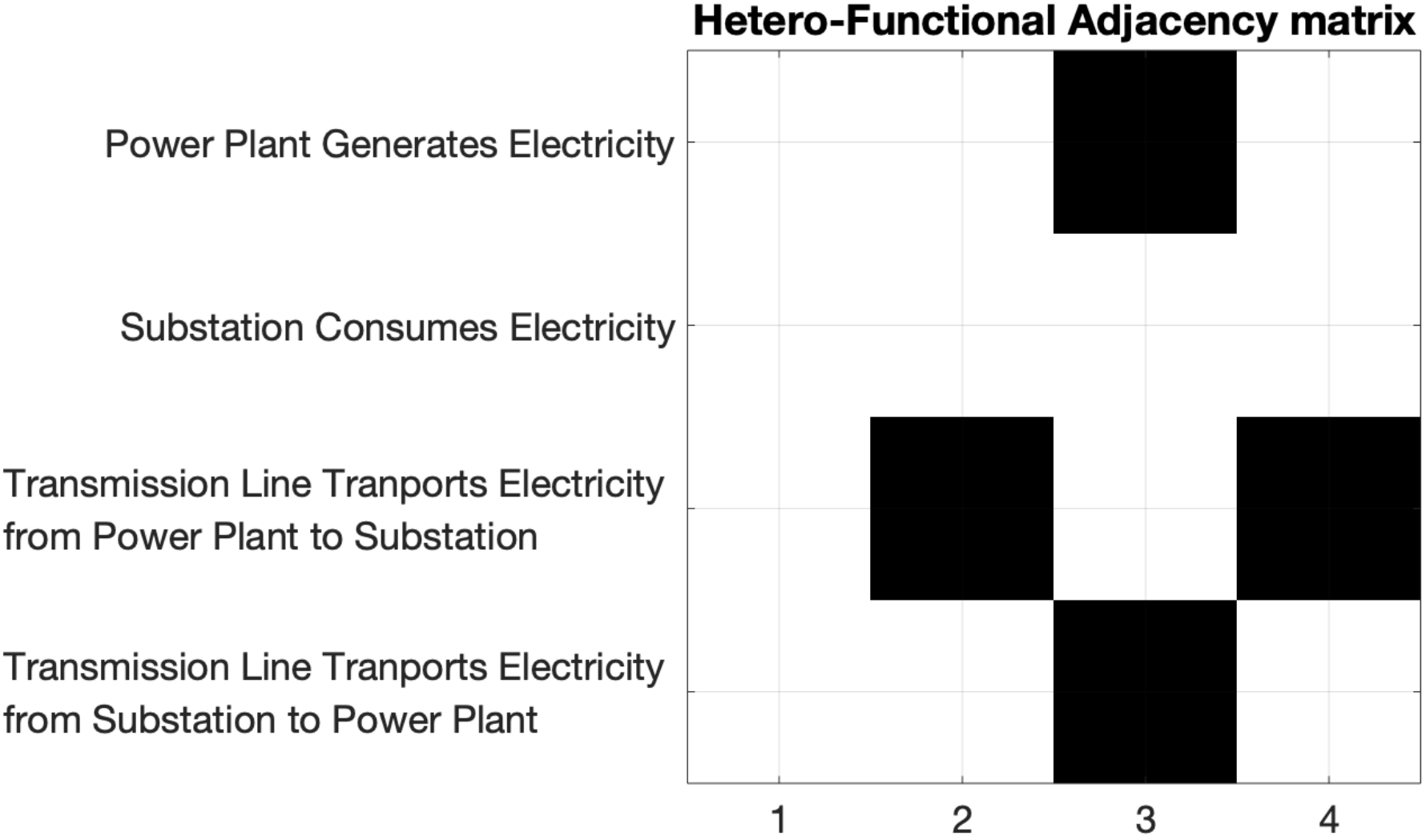}\label{ex_hfg}}
\caption{The example electric power system functional reference architecture \textbf{a} with a single instantiation of each system process pairing.  \textbf{b} gives the system knowledge base of the example system.  \textbf{c} gives the hetero-functional adjacency matrix of the example electric power system.}
\label{TG_vs_HF_C}
\end{figure}

The KML files extracted from the Platts Map Data Pro require several steps of data processing to produce a single XML file that is compatible with the hetero-functional graph-theory toolbox\citeM{Hegde:2020:ISC-JR03}.  
\begin{enumerate}
    \item First, any resources marked as canceled, retired, or shutdown are removed from the KML file.
    \item Second, any resources with duplicate GPS locations are merged such that only one resource exists per GPS location.  The meta-data for this resource is adopted from one of the other resources giving preference to power plants, then generation units and then substations as a last option.
    \item Third, the remaining individual generation units are classified as power plants.  These three steps yield a consistent set of formal nodes.
    \item Fourth, any transmission lines without a well-defined origin or destination formal node are removed.
    \item Fifth, the remaining resources are organized into a formal graph{\color{black}\citeM{leskovec:2016:00}}.  All isolated and sub-component nodes and edges not part of the largest connected component are identified.  These isolated nodes and clusters are removed from the system.
    \item Sixth, the system process for each type of resource is inferred.  Power plants ``generate electric power", substations ``consume electric power", and transmission lines ``transport electric power from origin to destination" and ``transport electric power from destination to origin."  Fig. {\color{black}\ref{ex_JS}} shows the knowledge base of an electric power system where the four types of system capabilities mentioned above are instantiated only once.  Fig. {\color{black}\ref{ex_hfg_act}} shows the associated SysML activity diagram{\color{black}\cite{Crawley:2015:00}}.  
    \item Lastly, the electric power system functional reference architecture shown in Fig. {\color{black}\ref{ex_hfg_act}} is encoded in the XML file as three valid pairs of system processes; 1.) transmission follows generation, 2.) transmission follows transmission, and 3.) consumption follows transmission.  Any other pairs of system processes are invalid and impose constraints in the system sequence constraints matrix $K_\rho$.  Fig. {\color{black}\ref{ex_hfg}} shows the hetero-functional graph.     
\end{enumerate}
    
The hetero-functional graph theory toolbox\citeM{Hegde:2020:ISC-JR03} is then used to produce the formal and hetero-functional graphs from the XML file described above.  The resulting formal graph is composed of $69,386$ formal nodes (i.e. power plants and substations) and $105,826$ formal edges (i.e. transmission lines).  As stated in Section A of the methods, these nodes and edges constitute  the $175,212$ resources in the system knowledge base while the system processes are defined as shown in Fig. {\color{black}\ref{ex_JS}}.  The resulting system knowledge base $J_S$ has a size of $4,814,416,999 \times 175,212$.   Once $J_S$ is formed, the hetero-functional adjacency matrix is formed using Equation \eqref{A_rho}.   $K_\rho$ is formed in a pairwise fashion observing the five types of constraints identified in Section A of the methods.   The first four types of constraints are checked numerically, while the last is drawn from the functional pairs in Step 6 above.  The resulting hetero-functional adjacency matrix $A_\rho$ has a size of $(8.435\times10^{14}) \times (8.435\times10^{14})$.  Equation {\color{black}\eqref{A_rho}} is then used to eliminate empty rows and columns.  The projected hetero-functional adjacency matrix $\tilde{A}_\rho$ of the American electric grid has a size $370,220 \times 370,220$ with $1,709,691$ capabilities as filled elements.  From this point, it was straightforward to calculate the cumulative degree distribution of AEPS' hetero-functional and formal graphs.    
\subsection{\textbf{Comparing the Attack Vulnerability of Scale-free Formal and Hetero-functional Graphs}}
This paper used nodal attacks to compare the attack vulnerability of formal graphs and hetero-functional graphs.  In nodal attacks, a set of nodes are identified either at random or by a targeting heuristic such as greatest degree centrality.  These nodes are subsequently removed from their respective graphs.  In addition to the identified nodes being removed, all edges connected to the removed nodes are also removed\citeM{Newman:2009:00}.  The same physical intuition was maintained between the hetero-functional graph and formal graph during attacks.  Rather than attacking a given system capability (as a node in a hetero-functional graph), an entire system resource (as a node in a formal graph) was attacked.  All of the associated system capabilities (i.e. the filled elements in the associated column of the knowledge base) were removed.   Because the formal edges had been removed in the case of the formal graph, their system capabilities were also removed.   For example,  if node $n_3$ in Fig. {\color{black}\ref{TG_vs_HFGT}} is attacked, then edges $e_1$, $e_3$, and $e_4$ are removed as well.  In the hetero-functional graph, first capabilities $\psi_3-\psi_6$ would be removed to reflect the loss of $n_3$ and then capabilities $\psi_8,\psi_{10},\psi_{11},\psi_{12}$ would all be removed to reflect the loss of edges $e_1$, $e_3$, and $e_4$. 

The nodal attacks were applied as both random and targeted attacks.  Rather attacking a single node at a time, for computational simplicity, each attack iteration removed one percent of the initial node count.  In the case of a random attack, the one percent of formal graph nodes were randomly selected and then removed.  The corresponding capabilities in the hetero-functional graph were then removed as well.  In the case of a targeted attack, the greatest degree centrality heuristic was used to identify the targeted nodes.  This one percent of nodes were then removed from the formal graph and the corresponding capabilities were then removed from the hetero-functional graph; thus maintaining the same physical intuition.  The most central nodes were then reevaluated and attacked again.  

Fig \ref{TG_vs_HF_cent_size}a was developed using a random attack and measuring the average degree distribution \citeM{Newman:2009:00} of each graph at each attack iteration.  Fig. \ref{TG_vs_HF_cent_size}b was developed using a random attack and measuring the component sizes of each graph at each attack iteration.  The largest weakly connected component size \citeM{Newman:2009:00} was measured and divided by the total remaining nodes in each graph for the largest cluster relative size.  The sizes of the remaining connected component (excluding the largest connected component) were then averaged together to get the isolated cluster average size at each attack iteration.  Fig \ref{TG_vs_HF_cent_size}c utilizes the same measures as Fig \ref{TG_vs_HF_cent_size}b.  However, a targeted attack based upon greatest degree centrality heuristic was applied to target central nodes.

\subsection{\textbf{Predicting the Structural Resilience of the AEPS' Migration to a Decarbonized System Architecture}}
In Fig \ref{add_cent_fg_hfg}, the resilience effects of adding the architectural improvements of distributed generation, meshed transmission, and storage were measured.  Distributed generation was added randomly to 30\% of all the substations; which amounted to $19,828$ substations gaining such capabilities.  When adding meshed distribution lines, each node with only a single connection to another node was connected to the nearest node to which it was not already connected.  An additional 20\% of transmission lines were thus added to the AEPS; which amounted to $16,953$ additional transmission lines.  Storage was added to every buffer that had the potential for generating electricity.  These nodes include power plants and substations that were designated to receive distributed generation capabilities.  Storage was thus added to 33\% of all buffers which amounted to $23,088$ additional energy storage resources.  

Each type of improvement was added to the AEPS over a series of 100 iterations.  The additions of distributed generation, meshed transmission, and storage were analyzed individually.  Fig. \ref{add_cent_fg_hfg}a and Fig. \ref{add_cent_fg_hfg}b measure the average degree centrality \citeM{Newman:2009:00} of the formal graph and hetero-functional graph respectively.  After each iteration of adding an architectural improvement to the formal graph and the corresponding capabilities to the hetero-functional graph, the average degree centrality was measured for both graphs.  It is notable that since both distributed generation and storage are functional additions, the formal graph sees no structural change.  Therefore, it sees  no measurable change in resilience from distributed generation or storage \cite{Thompson:2020:SPG-C68}.  However, because the nodes in the hetero-functional graph are system capabilities, there is a measurable structural change in the average degree centrality of the hetero-functional graph as distributed generation and storage are added.  Fig. \ref{add_cent_fg_hfg}c follows the same architectural improvements as Fig. \ref{add_cent_fg_hfg}a and \ref{add_cent_fg_hfg}b while measuring the hetero-functional graph's Latent Engineering Resilience (LER) \cite{Farid:2015:ISC-J19}.  The LER was normalized by the original base case AEPS with no architectural improvements.  The addition of distributed generation and storage report linear increases in the LER while the addition of meshed transmission lines results in an exponential increase in the LER.  Intuitively, as lines are added to the AEPS each step of a service delivery path has increasingly more path options; thus exponentially growing the number of deliverable service paths.

\subsection{\textbf{Assessing the Resilient Response of Several Decarbonized System Architectures of the AEPS to Random and Targeted Attacks}}

Fig \ref{HFG_LER_attack}a and Fig \ref{HFG_LER_attack}b measure the LER of the AEPS with different architectural improvements under random and targeted attacks respectfully.  Just as previous figures maintained an equivalent physical intuition when attacking hetero-functional and formal graphs that intuition was maintained in Fig \ref{HFG_LER_attack}.  Physical buffer resources (i.e. formal nodes) in corresponding formal graphs were selected for random or targeted attacks and all of the associated capabilities of the buffer and its connected edges were removed from the hetero-functional graph.  Attacks were applied to one percent of the formal graph nodes at a time in both random and targeted attacks.  By maintaining the same physical intuition, removing one percent of nodes being from a formal graph could result in removing more that one percent of the hetero-functional graph nodes.  This phenomena is especially noticeable in Fig. \ref{HFG_LER_attack}b as the initial targeted attack results in removing over five percent of the nodes from the hetero-functional graphs.  When applying targeted attacks in Fig. \ref{HFG_LER_attack}b, the greatest degree centrality heuristic was used to identify the targeted formal nodes.  

After each attack iteration, the LER of the remaining hetero-functional graph was re-evaluated.  In all cases, the LER was normalized by the initial unimproved base case AEPS.  Thus, the base case takes an initial value of 1 while the architecturally improved AEPSs measure an initial LER value greater than 1.  Under random attacks the LER gradually decreases following the regression $LER=\alpha(100-x)^2e^{\beta (100-x)}$.  However, the targeted attack causes a large initial drop paired with a faster decrease in the LER following the regression $LER=\alpha e^{\beta x}$.  The initial LER drop is caused because the attack of one percent of attacked formal nodes removes over five percent of the nodes in the hetero-functional graph.  Using degree centrality as the targeting method, buffers with the most transmission lines are removed first.  The first attack therefore removes power plants, substations, and the largest number of transmission lines in a single attack.  As buffers connected to the most transmission lines are targeted, there are increasingly fewer path options for services to be delivered.  Thus, similar to the addition of meshed transmission lines in Fig. \ref{add_cent_fg_hfg}c which yields an exponential increase in the LER, targeting buffers by degree centrality yields an exponential decay in the LER.

\subsection{\textbf{Data Availability}}
The data that support the findings of this study are commercially available from Platts but restrictions apply to the available dataset, which were used under license for the current study, and are not publicly available.  Data are however available from the authors upon reasonable request with express written consent of Platts.

\footnotesize{
\bibliographystyleM{IEEEtran}
\bibliographyM{0-AEPS_Resil}}

\section{\textbf{Author Contributions}}
Dakota Thompson was the lead author and primary analyst of the quantitative results.  
Wester C. H. Schoonenberg was the secondary analyst to the quantitative results.  Amro M. Farid designed the research scope and questions and guided conduct of the research to a conclusion.

\section{\textbf{Additional Information}}
Correspondence and requests for materials should be addressed to Dakota Thompson at Dakota.J.Thompson.Th@Dartmouth.edu.  Reprints and permissions information is available at www.nature.com/reprints.

\end{document}